\documentclass[nofootinbib,preprint,
showpacs,preprintnumbers,amsmath,amssymb]{revtex4}
\usepackage{latexsym}
\usepackage{hyperref}
\usepackage{epsfig, psfrag, graphicx}

\newcommand{\be}{\begin{equation}}
\newcommand{\ee}{\end{equation}}
\newcommand{\ben}{\begin{eqnarray*}}
\newcommand{\een}{\end{eqnarray*}}
\newcommand{\bea}{\begin{eqnarray}}
\newcommand{\eea}{\end{eqnarray}}
\newcommand{\bdm}{\begin{displaymath}}
\newcommand{\edm}{\end{displaymath}}
\newcommand{\ba}{\begin{align}}
\newcommand{\ea}{\end{align}}
\newcommand{\lb}{\label}

\begin{document}

\title{\bf Quantum phantom cosmology}

\author{Mariusz P. D\c{a}browski}
\email{mpdabfz@sus.univ.szczecin.pl}
\affiliation{Institute of Physics, University of Szczecin,
Wielkopolska 15, 70-451 Szczecin, Poland.}
\author{Claus Kiefer}
\email{kiefer@thp.uni-koeln.de}
\affiliation{ Institut f\"ur Theoretische Physik,
Universit\"at zu K\"oln, Z\"ulpicher
Str. 77, 50937 K\"oln, Germany.}
\author{Barbara Sandh\"ofer}
\email{bs324@thp.uni-koeln.de}
\affiliation{ Institut f\"ur Theoretische Physik,
Universit\"at zu K\"oln, Z\"ulpicher
Str. 77, 50937 K\"oln, Germany.}
\date{\today}
\begin{abstract}
We apply the formalism of quantum cosmology to models containing
a phantom field. Three models are discussed explicitly:
a toy model, a model with an exponential phantom
potential, and a model with phantom field accompanied by a negative
cosmological constant. In all these cases we calculate the classical
trajectories in configuration space and give solutions to the
Wheeler--DeWitt equation in quantum cosmology. 
In the cases of the toy model and the model with exponential potential
we are able to solve the Wheeler--DeWitt equation exactly.
For comparison,
we also give the corresponding solutions for an
ordinary scalar field. We discuss in particular
the behaviour of wave packets in minisuperspace. 
For the phantom field these packets
disperse in the region that corresponds to the Big Rip singularity.
This thus constitutes a genuine quantum region 
at large scales, described by a regular
solution of the Wheeler--DeWitt equation. 
For the ordinary scalar field, the Big-Bang singularity is avoided.
Some remarks
on the arrow of time in phantom models as well as on the relation of
phantom models to loop quantum cosmology are given.
\end{abstract}
\maketitle


\section{Introduction\label{intro}}

It is a striking fact that our universe is currently accelerating
\cite{supernovaeold}. A major open problem is to provide a fundamental type
of matter which may be responsible for this, since any of the forms
of matter we know from our experience cannot explain this phenomenon.
This matter is not visible,
but provides a dominant fraction of the energy density in
the universe and was therefore given the name `dark energy'
(for a current review see for example \cite{ed+shinji}).
The best known, and perhaps the simplest candidate for such a matter is
the cosmological constant,
but theoretical physics provides more options.
One of them is an evolving scalar field with appropriate kinetic and
potential energies. In general, it may mimic various types of
matter during different periods of the cosmological evolution.

Dark energy is characterized by negative pressure which
causes the repulsion of
matter in the universe and, as a consequence, its acceleration. In terms of the
standard energy conditions in general relativity \cite{he}, dark energy must
violate the strong energy condition $\rho + 3p > 0$, $\rho > 0$.
Assuming a barotropic equation of state of the matter in the
universe, $p=w\rho$ ($w=$ constant), where $p$ and $\rho$ are the pressure
and the density of dark energy, respectively, it requires that $w < - \frac{1}{3}$.

However, according to more recent observations
\cite{supernovaenew,riess2004},
dark energy is even more biased towards larger negative values
of the barotropic index $w\lesssim -1$. This means that it
would have to violate the null energy condition $\varrho + p >0$ \cite{he}
and, consequently, all the remaining energy conditions such as: the weak energy condition
$\varrho > 0, \varrho + p > 0$, and the dominant energy condition
$\varrho >0,  -\varrho < p < \varrho$. Dark energy of this type was dubbed
{\em phantom} \cite{Caldwell,phantom}.
A phantom may be represented by an evolving
scalar field which possesses negative kinetic energy (often called `ghost field').
Although phantom fields lead to various problems \cite{instab},
they are observationally supported
as a possible source of dark energy and deserve thorough investigation
(but see \cite{kaloper} for an alternative view).
Moreover, there may exist phantom fields without pathologic behaviour
in the ultraviolet regime \cite{Rubakov}.\\
Phantom models of the universe admit a new type of singularity
called a Big-Rip singularity \cite{Caldwell,phantom,Starobinsky}.
At the Big Rip, energy density and
pressure diverge as the scale factor $a(t)$ goes to infinity
at a finite time. This is different from an ordinary Big Crunch
singularity, which leads to a blow-up of the energy density and pressure as
the scale factor approaches zero at a finite time.
Another possible singularity is the Big Brake where the expansion rate
is zero and the acceleration rate approaches minus infinity \cite{GKMP}. 
Besides, more exotic types of singularities may appear 
such as the sudden future singularity \cite{new1}, 
the generalized sudden future singularity \cite{new2}
 where there is a blow-up of the higher-order derivatives 
of the scale factor with smooth evolution of the scale factor and the 
energy density, 
the type III singularity \cite{new3} and 
the type IV singularity \cite{new4}
 where the evolution of the scale factor is smooth. 
These singularities have weaker properties than a Big Rip \cite{new5}.

In Ref. \cite{marphan}, classical phantom cosmologies were
studied and a large variety of possible
cosmological scenarios were found. Also, the duality relation
between standard matter and phantom matter models was revealed
(see also \cite{lazkoz03,triality}) which has an analogon in the duality
symmetry present in superstring cosmology \cite{meissner}.

It is worth mentioning that, once the supernova data are analyzed
in a prior-free manner,
an evolving equation of state with a time-dependent barotropic index $w = w(t)$
for the dark energy is favoured (cf. \cite{ASSS,w(t),onemli}). Such
models were also studied in the quantum context in
\cite{bigtrip} where a canonical momentum was attached to a
time-dependent barotropic index. Some authors have also studied the
thermodynamical properties of phantom models
\cite{Diaz04,pavon05}.

In all these investigations, an evolving universe
was described by classical cosmology. Quantum effects were only
studied in certain phases of the evolution, for example, close to
a singularity, cf. \cite{NO}, without applying quantum theory to the
universe as a whole.
It is the purpose of this paper to accomplish the latter goal, that is,
to discuss quantum cosmology with
phantom fields. The interest in this is due to the fact that for both experimental and
theoretical reasons it seems that quantum theory is universally valid
\cite{deco}. Therefore, the universe as a whole has to be described
by quantum theory. If phantom fields play a dominant role, it has to be
investigated whether this causes deviations from the standard formalism
of quantum cosmology and whether there are interesting physical consequences.

Quantum cosmology must be based on a theory of quantum gravity \cite{OUP}.
Candidates for such a theory include string theory, loop quantum gravity
and quantum geometrodynamics. Our present analysis will, like most
investigations of quantum cosmology, be based on the Wheeler--DeWitt
equation of quantum geometrodynamics. Independently of the correct theory
of quantum gravity, this framework should yield an adequate description
at least on the energy scales below the Planck scale
(if not on all scales). If one approaches the
Planck scale, modifications such as loop quantum cosmology \cite{Bojowald}
might become necessary. The investigations in our paper are
independent of such modifications and will be discussed in a
future paper.

A central feature of the Wheeler--DeWitt equation is its local
hyperbolic signature \cite{OUP,Zeh}. In regions of configuration space
near closed Friedmann cosmologies, it is globally hyperbolic, that is, there
is only one minus sign in the kinetic term \cite{Giulini}. The negative
part of the kinetic term is related to the scale factor of the Friedmann
model, which in a certain sense thus plays itself the role of
a phantom field. The presence of an indefinite kinetic term is
intimately connected with the attractive nature of gravity \cite{GK}.

Besides its hyperbolic character,
the most important feature of the Wheeler--DeWitt equation
is its independence of an external time
parameter \cite{OUP,Zeh}. This holds, in fact, for every system that
is reparametrization invariant at the classical level. Consistent
discussions of quantum cosmology must thus be based on the
intrinsic structure of this equation and avoid the use of an intuitive
but wrong picture of an external Newtonian time. For this purpose it is
necessary to study the classical trajectories in
a {\em configuration space} where the classical time parameter $t$ is eliminated.

The structure of the Wheeler--DeWitt equation is important
for the imposition of boundary conditions in quantum cosmology.
In the hyperbolic case one has a wave equation whose form suggests
imposing boundary conditions at constant values of the scale factor,
$a$. This is of importance, for example, if one attempts to
construct wave packets that follow the classical trajectories in
configuration space like standing tubes \cite{Ki88,Ki90,GS01}.
It is also crucial for an understanding of what pre- and post big
bang phases mean in quantum string cosmology \cite{DK,QSC}.
The origin of the arrow of time can in principle be
traced back to the structure of this wave equation \cite{Zeh,CZ,KieZeh}.

The presence of a phantom field changes the structure of the
Wheeler--DeWitt equation: If only the phantom is present besides
the scale factor (`phantom dominance'), its structure
becomes elliptic, while in the general case it becomes of a mixed
(ultrahyperbolic) nature.
This has implications for the imposition of boundary conditions.
A change of signature in the Wheeler--DeWitt equation has hitherto been
noticed in the presence of non-minimally coupled fields \cite{Ki89}.
In our paper we shall present the formalism of quantum phantom cosmology
and some of its main physical consequences.

Our paper is organized as follows. In Sec. \ref{Classical} we shall study
and solve the classical equations of motion for the phantom field
in a Friedmann universe. After the presentation of the necessary
equations, we give the solutions for the classical trajectories in
configuration space for three models: A toy model with
vanishing phantom potential (Sec. \ref{class_toymodel}), a model
with exponential phantom potential (Sec. \ref{class_nolambda}), and a
model with cosh-potential and a negative cosmological constant
(Sec. \ref{neglambda}). For comparison, in all these cases, we give the results for
a non-phantom scalar field. Sec. \ref{QPC} contains in the same order
the discussion of the quantum theory for these models, both for
a phantom field and a corresponding ordinary scalar field.
We are able to solve the Wheeler--DeWitt equation exactly for the
toy model and the model with exponential potential.
In particular, we discuss wave packet solutions and find that
quantum effects dominate in the region of the classical Big-Rip singularity.
Therefore,  {\it quantum effects
occur at large scales}. Since the solutions of the
Wheeler--DeWitt equation are regular there, the Big-Rip singularity
has vanished in the quantum theory. Furthermore, in the realistic scalar field
models, the wave function vanishes at the Big Bang. Thus, this
singularity is likewise excluded in the quantum theory.
In Sec. \ref{Sect4} we give a
summary of the results and the outlook of the problem
of the arrow of time and possible modifications of the obtained picture
due to loop quantum cosmology.


\section{Classical phantom cosmologies in configuration space\label{Classical}}

\subsection{Classical equations}

We consider a Friedmann universe with scale factor $a(t)$ and
a homogeneous scalar field $\phi(t)$. We assume here that the $\phi$-field
dominates over other matter degrees of freedom, so that it
is the only degree of freedom besides the scale factor.
The action reads
\begin{equation}
\lb{action}
S = \frac{3}{\kappa^2}\int\mathrm{d}t\ N\left(-\frac{a\dot{a}^2}{N^2}
+{\mathcal K}a-\frac{\Lambda a^3}{3}\right)
+\frac{1}{2}\int\mathrm{d}t\ Na^3\left(\ell\frac{\dot{\phi}^2}
{N^2}-2V(\phi)\right)\ .
\end{equation}
Here, $\kappa^2=8\pi G$, $N$ is the lapse function,
$\Lambda$ is the cosmological constant, $V(\phi)$ is a
potential of the field $\phi$,
${\mathcal K}=0,\pm1$ is the curvature index, and we have set $c=1$.
The parameter $\ell$ distinguishes
between a phantom field (where $\ell=-1$) and an ordinary
scalar field (where $\ell=+1$).

We set $N=1$, so the time parameter is the standard Friedmann cosmic time.
The action then becomes
\be
\lb{action2}
S=\frac{3}{\kappa^2}\int{\mathrm d}t\ (-a\dot{a}^2+{\mathcal K}a-\frac{\Lambda}{3} a^3)
+\frac12\int{\mathrm d}t\ (a^3\ell\dot{\phi}^2-2a^3V(\phi))\ .
\ee
The canonical momenta are given by
\be
\lb{momenta}
\pi_a=-\frac{6a\dot{a}}{\kappa^2}\ , \quad \pi_{\phi}=\ell a^3\dot{\phi}
\ .
\ee
The canonical Hamiltonian ${\mathcal H}$, which is constrained to
vanish, reads
\be
\lb{constraint}
{\mathcal H}=-\frac{\kappa^2}{12a}\pi_a^2+\frac{\ell}{2}\frac{\pi_{\phi}^2}
{a^3}+a^3\frac{\Lambda}{\kappa^2} + a^3V-\frac{3{\mathcal K}a}{\kappa^2}=0\ .
\ee
Expressed in terms of the `velocities', see \eqref{momenta},
this constraint becomes identical to the Friedmann equation,
\be
\lb{Friedmann}
\left(\frac{\dot{a}}{a}\right)^2\equiv H^2=
\frac{\kappa^2}{3}\left(\ell\frac{\dot{\phi}^2}{2}+V(\phi)\right)
+ \frac{\Lambda}{3}-\frac{\mathcal K}{a^2}\ .
\ee
The term in parentheses is the energy density of the scalar field,
\be
\label{rho}
\rho\equiv \ell\frac{\dot{\phi}^2}{2}+V(\phi)\ .
\ee
We recognize that for the standard scalar field $(\ell=1)$,
no classically forbidden regions
in configuration space exist due to the indefiniteness
of the total kinetic term.
This is different from the phantom case
$(\ell=-1)$, where only the region
\be
\lb{classregions}
V(\phi)+ \frac{\Lambda}{\kappa^2}-\frac{3{\mathcal K}}{\kappa^2a^2} \geq 0
\ee
is classically allowed (this restriction is due to the negative
definiteness of the total kinetic term).

The field $\phi$ obeys the second-order equation of motion
\be
\lb{phi}
\ddot{\phi}+3\frac{\dot{a}}{a}\dot{\phi}+\ell V'(\phi)=0\ ,
\ee
which is equivalent to the conservation equation $\dot{\rho}
+3H(\rho+p)=0$, provided the standard perfect-fluid energy-momentum tensor is
introduced. This equation is trivially fulfilled by the cosmological constant
$\Lambda$ ($\rho=$ constant, $p=-\rho$).
In \eqref{phi} we recognize a formal reversal of the potential in
the phantom case compared to an ordinary scalar field case, since
the sign in front of the $V'$-term changes.
With the help of \eqref{Friedmann}, the second-order equation for $a$ can
be put into the form
\be
\lb{ddota}
\frac{\ddot{a}}{a}-\frac{\Lambda}{3} + \frac{\kappa^2}{3}\left(\ell\dot{\phi}^2
-V(\phi)\right)=0\ .
\ee
Again, assuming the perfect-fluid energy--momentum tensor, the scalar field
exerts the pressure
\be
\label{pe}
p\equiv \ell\frac{\dot{\phi}^2}{2}-V(\phi)\ .
\ee
Note that the case of a cosmological constant
is included by having the additional equation of state
\be
\label{Lambda}
p_{\Lambda} = - \rho_{\Lambda} = - \frac{\Lambda}{\kappa^2}\ .
\ee
Assuming a constant barotropic index $w$ for the
scalar field, we can use \eqref{rho} and
\eqref{pe} to find the relation between the scalar field
and its potential \cite{david},
\be
\label{virial}
V(\phi(t)) = \frac{\ell}{2} \frac{1-w}{1+w}
\dot{\phi}^2(t)\ , \hspace{0.3cm} w \neq -1\ .
\ee
This is analogous to the virial theorem in which the kinetic energy is
proportional to the potential energy of the field. However,
as has already been mentioned above, it may be more physical
to assume a time-dependent barotropic index \cite{ASSS,w(t),onemli}.

\subsection{Classical phantom trajectory
for vanishing phantom potential and vanishing cosmological constant\label{class_toymodel}}

In this section we shall consider a
simple model with field potential $V(\phi) = 0$ and cosmological constant
$\Lambda=0$. This leads to an equation of state
for stiff matter, $p=\rho$, $w=1$, in contrast to the current observational status
\cite{supernovaeold,supernovaenew}. However, such an evolution may
perhaps be valid in
ekpyrotic/cyclic scenarios where this matter dominates the
collapsing phase of the cosmological evolution \cite{turok}.
Moreover, in such a case the energy density $\rho<0$, and thus this model does not seem to represent dark energy
which is usually assumed to have positive energy density. However,
it captures interesting `phantom features', since it violates all
energy conditions, and it has the merit that it is easily manageable.
More realistic models will be discussed below.

In order to get classical solutions in the phantom case $(\ell=-1)$
we have to choose ${\mathcal K}=-1$ in \eqref{classregions}.
Since we are interested in constructing wave packets from
the Wheeler--DeWitt equation, we want to find a classical
trajectory in {\em configuration space}, where the
classical time $t$ is eliminated. This is motivated by the fact that
no such parameter is present in the Wheeler--DeWitt equation.

Since $\phi$ is a cyclic variable, $\pi_{\phi}$ is constant, so
from \eqref{momenta} one has $\dot{\phi}^2=C_p^2/a^6$ with a
constant $C_p$. From \eqref{Friedmann} one then has (we choose $C_p>0$)
\be
\frac{{\mathrm d}\phi}{{\mathrm d}a}=
\pm\frac{C_p}{a\sqrt{a^4-\frac{\kappa^2C_p^2}{6}}}\ ,
\ee
which can easily be integrated to yield
\be
\phi(a)=\pm\frac{1}{\kappa}\sqrt{\frac32}{\mathrm{arccos}}
\frac{\kappa C_p}{\sqrt{6}a^2}\  .
\ee
For convenience, we choose $\kappa^2=6$. Then the solution reads
\be
\lb{phia1}
\phi(a)=\pm\frac12{\mathrm{arccos}}\frac{C_p}{a^2}\ .
\ee
The classical trajectory \eqref{phia1} has a minimum value of the scale factor,
$a_{\mathrm{min}}=\sqrt{C_p}$, and reaches infinite values of $a$
at finite values of $\phi=\pm \pi/4$. In this sense it resembles
a Big-Rip solution. However, with respect to $t$ the scale factor
reaches infinity only at $t=\pm\infty$ and, moreover, $\rho\propto
a^{-6}$ which is the density scaling appropriate to a stiff fluid.
Nonetheless, in configuration space the trajectory has some features
of a Big-Rip, and this is why this toy model is of interest.

For an ordinary scalar field $(\ell=1)$ and for ${\mathcal K=-1}$,
one gets instead of \eqref{phia1},
\be
\lb{phia2}
\phi(a)=\pm\frac12{\mathrm{arcsinh}}\frac{C_f}{a^2}\ .
\ee
There is no turning point; Eq. \eqref{phia2} just
describes two branches for which
$a\to\infty$ if $\phi\to 0$, and $a \to 0$ if $\phi\to \pm\infty$.
The two solutions \eqref{phia1} and \eqref{phia2} are depicted in
Figure~1.
For ${\mathcal K}=1$ one obtains the solution with a turning point
(arccosh instead of arcsinh) that was discussed in \cite{Ki88}.

\begin{figure}[bt]
\unitlength1cm
\psfrag{X}{\mbox{$a$}}
\psfrag{Y}{\mbox{$\phi$}}
\psfrag{Branch_+}{\mbox{$\phi_+$}}
\psfrag{Branch_-}{\mbox{$\phi_-$}}
\begin{center}
\scalebox{0.5}{\includegraphics[angle=0]{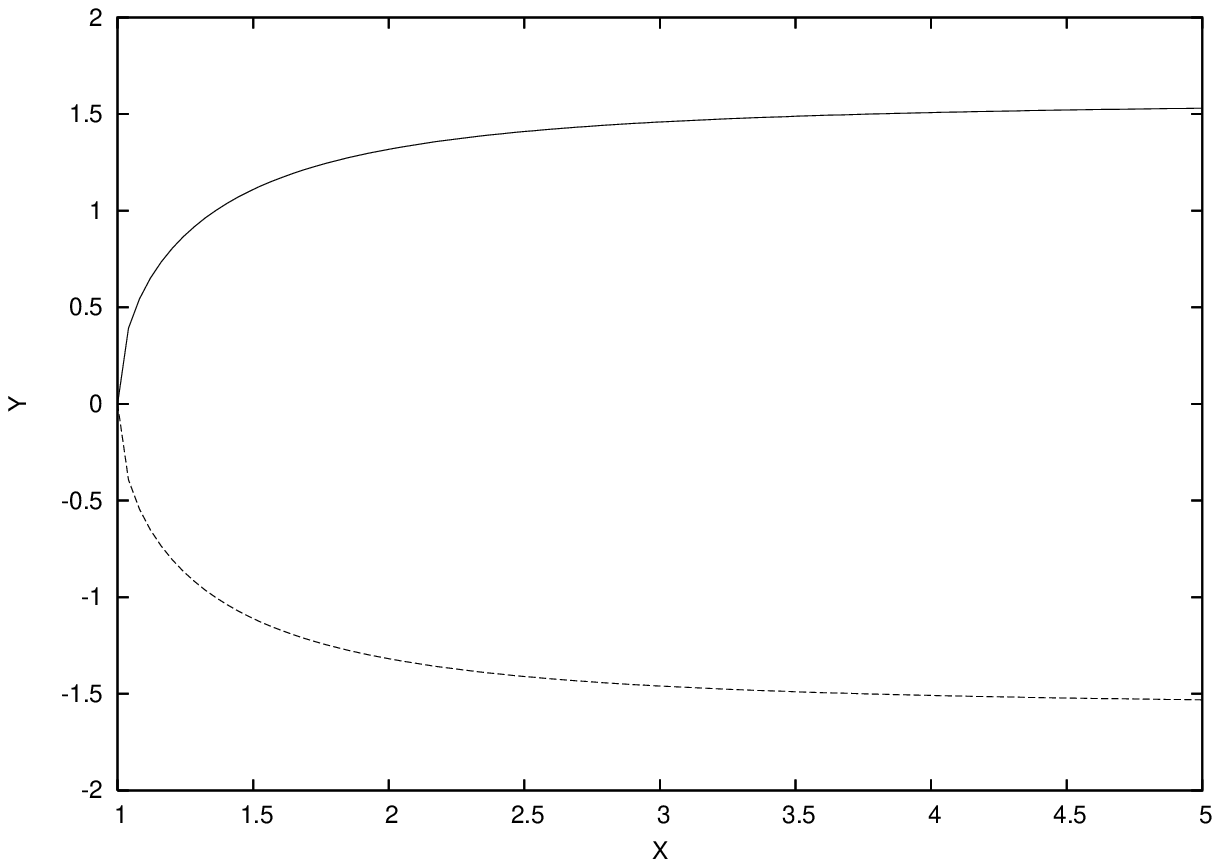}\hspace{50mm}\includegraphics[angle=0]{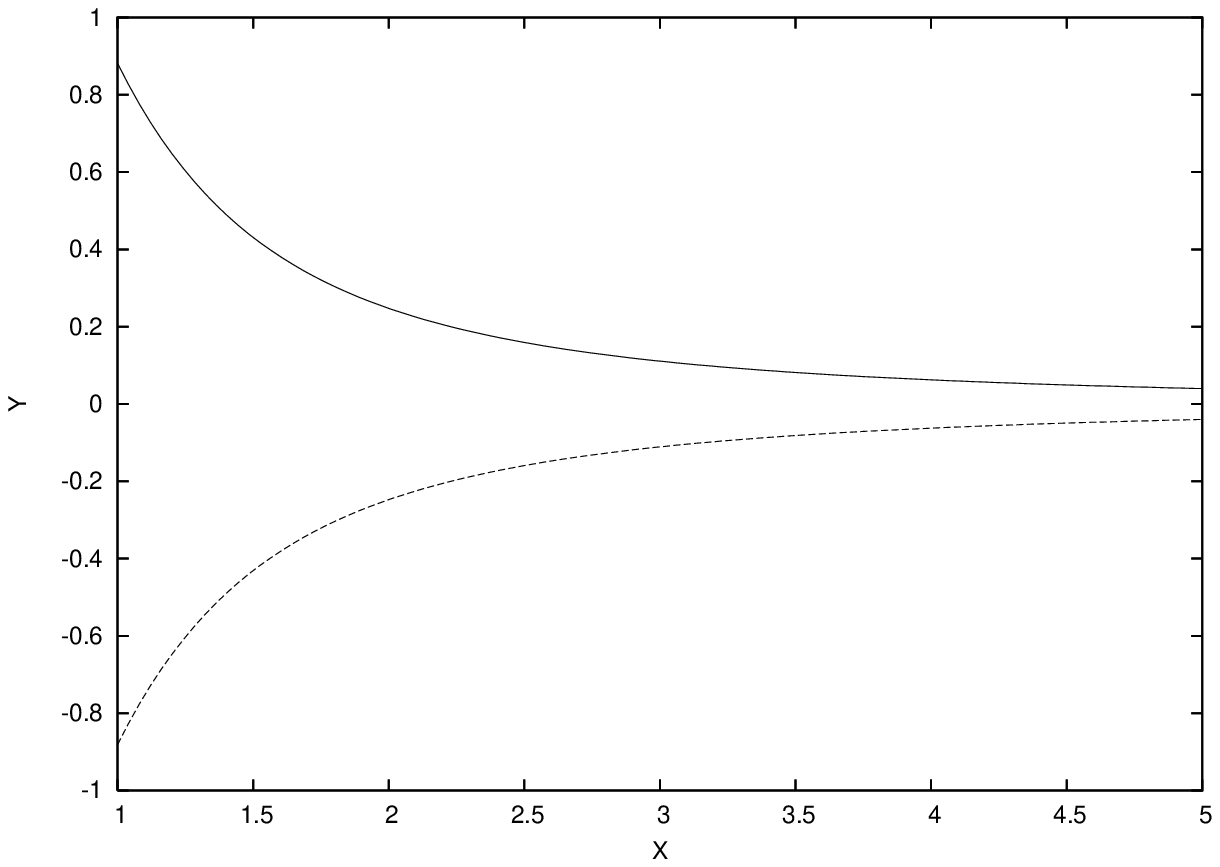}}
\caption{The classical trajectory in configuration space for the toy
  model with vanishing scalar field potential and vanishing cosmological
  constant. The diagram on the left-hand side shows the trajectory for the
  phantom field model. On the right-hand side the trajectory for the
  $\ell=1$ scalar field model is plotted.}
\end{center}
\end{figure}


\subsection{Classical trajectories for exponential scalar field potential and vanishing cosmological constant\label{class_nolambda} }

A model with phantom equation of state and a true Big-Rip singularity for
the phantom model appears if the potential in \eqref{action2} is chosen
to be exponential \cite{HaoLi1,HaoLi2} 
\be
\lb{exppot}
V\left(\phi\right)=V_0e^{-\lambda\kappa\phi}\ ,
\ee
and $\Lambda=0$.
Interest in this type of scalar field potentials in cosmology arose when it
became clear that the classical model has an attractor solution with scalar field
domination \cite{Copeland,Ratra}. This alleviates the fine-tuning problem of the
initial energy of the scalar field \cite{FerrJoyce}. Such an attractor exists
not only in the case of a conventional scalar field, but also for the phantom field
\cite{HaoLi1,HaoLi2}.
Exponential potentials for scalar fields arise in the context of
Kaluza--Klein theories \cite{KK1,KK2}, higher-derivative gravity in $\left(D+4\right)$
dimensions \cite{HDgrav1,HDgrav2}, higher-order gravity \cite{HOgrav},
supergravity
and superstring theories \cite{Halliwell,Yokoyama}, see also
\cite{FerrJoyce} for an overview.

In the following, we shall consider the case of a flat universe, $\mathcal
K=0$.
 From equation \eqref{classregions} one sees
immediately that for this choice of parameters neither the ordinary scalar field nor the phantom
field model possesses classically forbidden regions.
The classical equations of motion \eqref{phi} and \eqref{ddota} can be
transformed into a dynamical system with the Friedmann equation
\eqref{Friedmann}
as a constraint \cite{Copeland,HaoLi1,FerrJoyce,HaoLi2}. For
$\ell=-1$ and arbitrary values of $\lambda$, as well as for $\ell=+1$ and $\lambda<\sqrt6$, this system has
an attractor solution given by \cite{lazkoz03}:
\be
\lb{phi_sol}
\phi\left(t\right)=\frac{2}{\lambda\kappa}
\ln{\left[1+\ell\frac{ \lambda^2H_0}{2}\left(t-t_0\right)\right]}\ ,
\ee

\be
\lb{a_sol}
\frac{a}{a_0}=\left[1+\ell\frac{ \lambda^2H_0}{2}\left(t-t_0\right)\right]^{\ell\frac{2}{\lambda^2}}\ .
\ee
Introducing $\alpha\equiv\mathrm{ln}(a)$ for later convenience,
one obtains the following simple trajectory in configuration space,
\be
\lb{phi_alpha}
\phi\left(\alpha\right)=\ell\frac{\lambda}{\kappa}\alpha\ .
\ee

For this attractor solution, the `kinetic energy' defined from
\eqref{Friedmann} --- writing this equation in the form
$\ell E_{\rm kin}+E_{\rm pot}=1$ --- is given by
(using $\kappa^2=6$ in the second step)
\bdm
E_{\rm kin}\equiv\frac{\kappa^2}{6}\left(\frac{d\phi}{d\alpha}\right)^2
=\frac{\lambda^2}{6}
\edm
and thus constant. Therefore, also the `potential energy' of the scalar field
is constant,
\bdm
E_{\rm pot}\equiv\frac{\kappa^2V}{3H^2}=
1-\frac{\ell\lambda^2}{6}\ .
\edm
The equation of state parameter of the field is
\be
w=-1+\ell\frac{\lambda^2}{3} \ .
\ee
Thus for $\ell=-1$, $\phi$ indeed describes a phantom field with $w<-1$, whereas
the scalar field with $\ell=1$ covers the range $w>-1$. Accordingly, the energy
density scales as $\rho=\rho_0\left(\frac{a}{a_0}\right)^{-\ell\lambda^2}$.
As expected, this yields a Big-Rip singularity for $\ell=-1$, 
since in the limit
$t\to t_1\equiv t_0-2\ell/{(\lambda^2H_0)}$ the energy density and the scale factor diverge.
For $t\to\infty$, $a$ and $\rho$ vanish.
This is in contrast to the  $\ell=1$ model: In the limit $t\to t_1$,
$a$ goes to zero and $\rho$ diverges, yielding a Big Bang, whilst for $t\to\infty$, $a$ diverges
and $\rho$ goes to zero.

\subsection{Classical trajectories for scalar field fluid and negative
  cosmological constant\label{neglambda}}

It is easy to obtain a simple set of classical
solutions for cosmological models with a negative
cosmological constant \cite{marphan}. In contrast to
a positive cosmological constant which supports cosmological
repulsion, the negative cosmological constant is a source of
attraction and can overcome the influence of repulsion from
dark energy with negative pressure such as cosmic strings,
domain walls, and phantom, see for example \cite{marphan}.
This allows models with a negative cosmological constant and
other fluids to evolve symmetrically between two singularities
with an extremum in between. In particular, {\em it is possible to have
an evolution beetween the two Big Rips which appear at finite cosmic time}, as
will be shown below.

We assume a flat universe, ${\cal K}=0$, with a negative
cosmological constant $\Lambda <0$ and a cosmological fluid
with barotropic equation of state $p=w \rho$; the latter will be mimicked by
a scalar field $\phi$ of either standard or phantom type.
In this case, the energy conservation equation gives
\be
\rho = C a^{-3(w+1)}\ ,
\ee
and the energy conservation equation \eqref{Lambda} for the cosmological
constant remains valid.
This can be used to solve the system \eqref{Friedmann} and \eqref{ddota}
in terms of the scale factor as
\be
\label{aD}
a(t) =  \left[A\sin{ \left(\frac{\vert D \vert}{\sqrt{3}}
( -\Lambda )^{\frac{1}{2}} t\right)}\right]^{\frac{1}{D}}\ ,
\ee
where
\be
\label{D}
D = \frac{3}{2}(1+w)\ , \hspace{0.3cm} \frac{6C}{A^2} = -\Lambda >
0 \ .
\ee
Using \eqref{D}, we can rewrite \eqref{virial} in the form
\be
\label{virial1}
V(\phi(t)) = \frac{\ell}{2} \frac{3-D}{D} \dot{\phi}^2(t)\ ,
\ee
which allows to write \eqref{rho} as
(note that as in Sec.~\ref{class_toymodel} we have assumed
$\kappa^2=6$),
\be
\label{aphi}
\rho =  \frac{3\ell}{2D} \dot{\phi}^2\ =
\frac{\dot{a}^2}{2a^2} - \frac{\Lambda}{6}\ .
\ee
With all these assumptions we are able
to calculate the evolution of the scalar field as
\be
\label{philam}
\phi(t) = \pm \frac{1}{\sqrt{3}\vert D\vert} \sqrt{\frac{D}{\ell}}
\ln{\vert \tan{\left(\frac{\vert D \vert}{2\sqrt{3}}( -\Lambda )^{\frac{1}{2}}
t\right)} \vert} \ .
\ee
Let us note that $\ell=+1, D>0$ for an ordinary scalar field, while
$\ell=-1, D<0$ for the phantom. Then, the above expressions make sense since
$D/\ell=\vert D\vert >0$.
For $D>0$ (negative cosmological term plus $w>-1$ fluid), the evolution
of the universe based on \eqref{aD} begins with a Big Bang at $t=0$, reaches a
maximum $a_{\rm max} = A^{1/D}$, and terminates with a Big Crunch at $t=\pi$.
For $D=-\vert D \vert <0$ (the phantom case),
 the evolution starts with a Big Rip at
$t=0$, reaches a minimum $a_{\rm min} = A^{-1/|D|}$, and terminates
with a Big Rip at $t=\pi$. The latter case is of special interest
because it allows a symmetric evolution of the scale factor
in the presence of a phantom field. This model may also be of interest
to study the cosmological arrow of time, see Sec.~IV.

A similar type of symmetric evolution appears in configuration
space. Using \eqref{aD} and \eqref{philam} to eliminate the classical
time coordinate, we obtain the trajectory
\be
\label{traj1}
\phi(a) = \pm \frac{1}{\sqrt{3}\vert D \vert} \sqrt{\frac{D}{\ell}}
\ln{ \left( \frac{a^D}{A + \sqrt{A^2 - a^{2D}}} \right) }\ .
\ee

\begin{figure}[bt]
\unitlength1cm
\psfrag{X}{\mbox{$a$}}
\psfrag{Y}{\mbox{$\phi$}}
\psfrag{Branch_+}{\mbox{$\phi_+$}}
\psfrag{Branch_-}{\mbox{$\phi_-$}}
\begin{center}
\scalebox{0.5}{\includegraphics[angle=0]{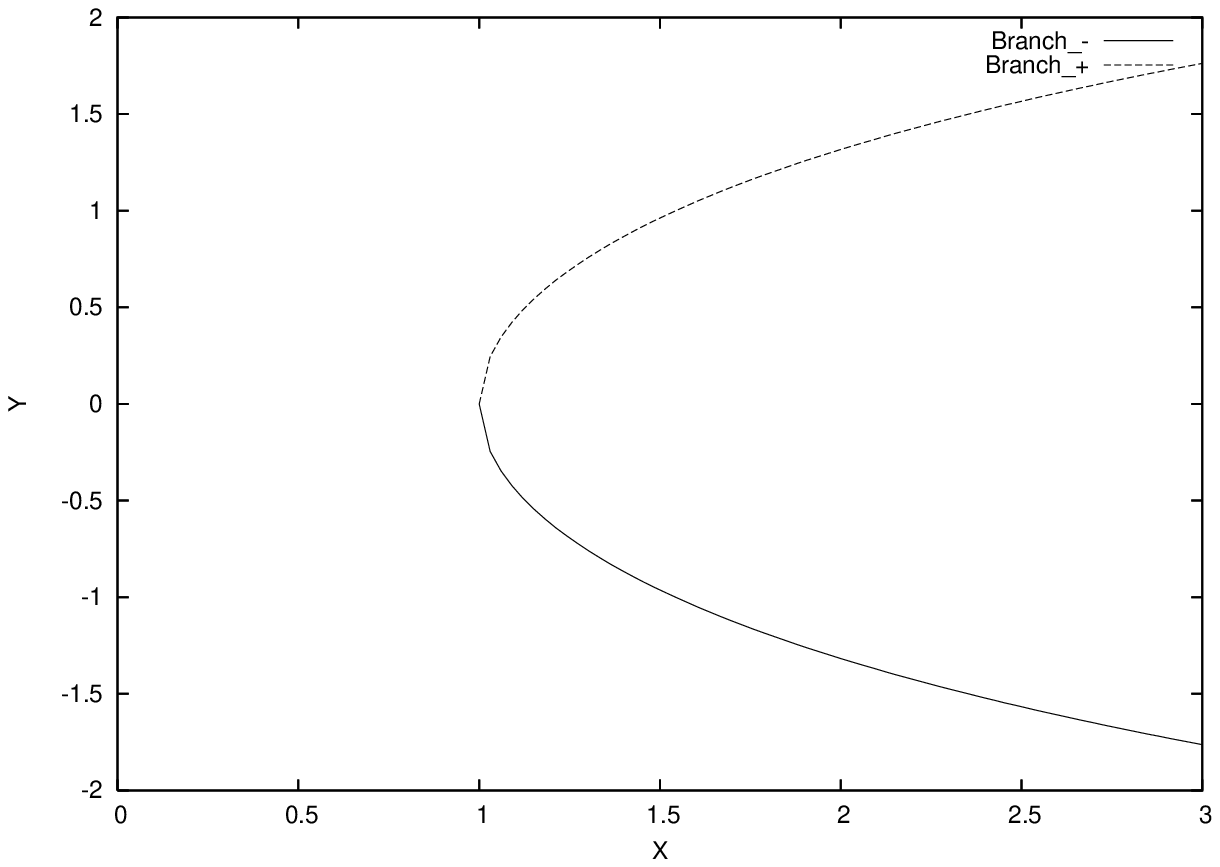}\hspace{50mm}\includegraphics[angle=0]{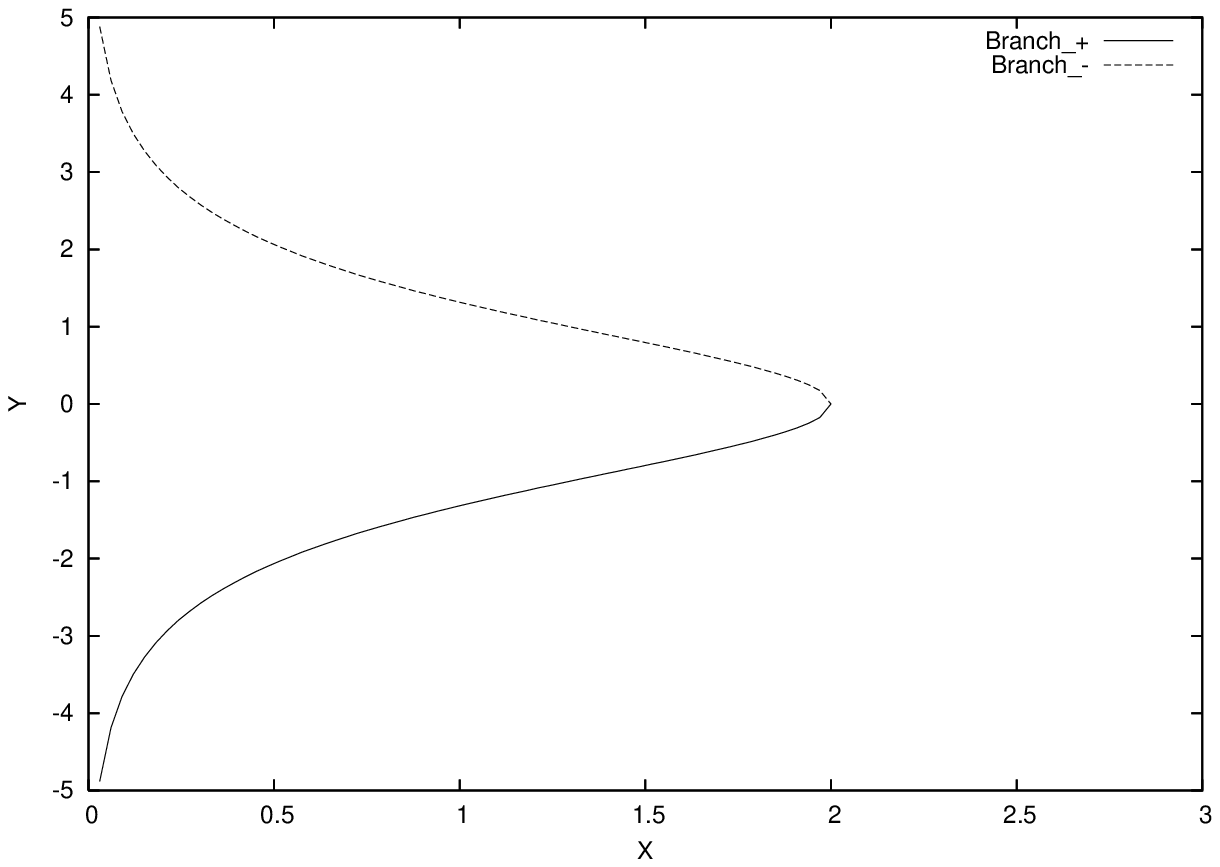}}
\caption{The classical trajectories in configuration space for the models
  with cosh-potential and negative cosmological constant. On the
  left-hand side,
  the trajectory for the phantom field model is shown. The classical
  trajectory  for the scalar field model $\ell=1$ is shown on the
  right-hand side. The similarity to the classical
  trajectories in the toy model in Sec. \ref{class_toymodel} is obvious.}
\end{center}
\end{figure}

From this we can see that there are two branches. For $\ell =-1$ each of them
extends to infinity, that is, $\phi \to \pm \infty$, for $a \to \infty$ and
reaches a minimum $\phi(a) = 0$, for $a_{\rm min} = A^{-1/\vert D\vert}$.
For $\ell=1$ one recognizes the presence of the maximum
$a_{\rm max}$. The trajectories in configuration space are depicted in
Figure~2.
 From \eqref{philam} and \eqref{aphi} one can reconstruct the
potential of the scalar field,
\be
\lb{cosh}
V(\phi)=V_0\cosh^2\left(\frac{\phi}{F}\right)\ ,
\ee
where
\bdm
V_0=-\frac{\Lambda}{3}\frac{\ell}{2}\left(3-D\right)F^2D\ , \quad
F=\frac{1}{\sqrt{3}\vert D \vert} \sqrt{\frac{D}{\ell}}\ .
\edm
Note that for $\ell=1$ the potential is positive only for $D<3$
(i.e., $w<1$). This
restriction is similar to the restriction $\lambda<\sqrt{6}$ in Sec.~\ref{class_nolambda}.

\section{Quantum cosmology for phantom and ordinary field\label{QPC}}

\subsection{Wheeler--DeWitt equation and phantom duality}

Quantization of the Hamiltonian constraint \eqref{constraint}
leads to the Wheeler--DeWitt equation. Choosing the Laplace--Beltrami
factor ordering and again the convention $\kappa^2=6$, it reads
\be
\lb{wdw1}
\left(\frac{\hbar^2}{2}a\frac{\partial}{\partial a}
a\frac{\partial}{\partial a}-\ell\frac{\hbar^2}{2}\frac{\partial^2}
{\partial\phi^2}+a^6\left(V(\phi)+\frac{\Lambda}{6}\right)-\frac{{\mathcal K}a^4}{2}\right)
\psi(a,\phi)=0\ .
\ee
Let us note that under the phantom duality \cite{lazkoz03}
\bea
a & \to & \frac{1}{\bar{a}}\ ,\\
\phi & \to & -i \bar{\phi}\ ,
\eea
for ${\mathcal K} = 0$ 
the Wheeler--DeWitt equation for $a, \phi$,
\be
\lb{wdw11}
\left(\frac{\hbar^2}{2}a\frac{\partial}{\partial a}
a\frac{\partial}{\partial a}-\ell\frac{\hbar^2}{2}\frac{\partial^2}
{\partial\phi^2}+a^6\left(V\left(\phi\right)+\frac{\Lambda}{6}\right)\right)
\psi(a,\phi)=0\ ,
\ee
transforms into the Wheeler--DeWitt equation for
$\bar{a},\bar{\phi}$, that is,
\be
\lb{wdw12}
\left(\frac{\hbar^2}{2}\bar{a}\frac{\partial}{\partial \bar{a}}
\bar{a}\frac{\partial}{\partial \bar{a}}+\ell\frac{\hbar^2}{2}\frac{\partial^2}
{\partial{\bar\phi}^2}+\frac{1}{\bar{a}^6}\left(V\left(i\bar\phi\right)+\frac{\Lambda}{6}\right)\right)
\psi(\bar{a},\bar{\phi})=0\ .
\ee
The transformation for $\phi$ is thus just a Wick rotation.

On the other hand, Eq. \eqref{wdw11} can conveniently be rewritten
in terms of the scale factor $\alpha \equiv \ln{(a)}$ as
\be
\lb{wdw2}
\left(\frac{\hbar^2}{2}\frac{\partial^2}{\partial\alpha^2}-
\ell\frac{\hbar^2}{2}\frac{\partial^2}{\partial\phi^2}+
e^{6\alpha}\left(V\left(\phi\right)+\frac{\Lambda}{6}\right)\right)\Psi(\alpha,\phi)=0.
\ee
It is this form of the Wheeler--DeWitt equation with which we shall
work in the following.
\subsection{Quantum phantom cosmology -- no phantom potential}

For vanishing potential $V=0$, $\Lambda=0$,
and ${\mathcal K}=-1$ the solution to the
phantom ($\ell=-1$) Wheeler--DeWitt equation \eqref{wdw1}
is found by a separation ansatz,
\be
\lb{psiaphi}
\psi_k(a,\phi)=C_k(a)\varphi_k(\phi)\ .
\ee
We choose
\be
\varphi_k(\phi)=e^{-ik\phi/\hbar}\ ,
\ee
because real exponentials would lead to exponentially increasing
wave functions for $\phi\to\pm\infty$ that would not
reflect classical behaviour. From \eqref{wdw1} one then gets the
following equation for the $C_k$ (primes denote derivatives with
respect to $a$),
\be
\lb{Ck}
a^2C_k''+aC_k'+\frac{1}{\hbar^2}(a^4-k^2)C_k=0\ .
\ee
Solutions of this equation are Bessel functions $Z_{k/2\hbar}(a^2/2\hbar)$.
However, we have to impose the boundary condition that
$\psi(a,\phi)\stackrel{a\to0}{\longrightarrow}0$ in order to
reflect the behaviour of the classical trajectories \eqref{phia1}
which have in configuration space a minimum with respect to $a$.
We therefore have to choose the Bessel function
$J_{k/2\hbar}(a^2/2\hbar)$ with $k>0$.

The connection to the classical solution \eqref{phia1}
should be performed through a formal WKB limit `$\hbar \to 0$'.
We thus have to look for an asymptotic expansion of $J$
where both the argument and the index are large.
We use the expression \cite{AS1}
\be
\lb{Bessel}
J_{\nu}(\nu z)=\left(\frac{4\zeta}{1-z^2}\right)^{1/4}
\left(\frac{\mathrm{Ai}(\nu^{2/3}\zeta)}{\nu^{1/3}}
+\frac{\exp(-\frac23\nu\zeta^{3/2})}{1+\nu^{1/6}\vert\zeta\vert^{1/4}}
{\mathcal O}\left(\frac{1}{\nu^{4/3}}\right)\right)
\ee
and set $\nu=k/2\hbar$, $z=a^2/k$. The choice for $\zeta$
depends on whether $z^2\geq1$ or $z^2<1$. Let us consider first
the case $z^2\geq1$ which corresponds to $a^4/k^2\geq1$. From
\cite{AS1} one sees that then
\be
-\zeta=\left(\frac32\sqrt{\frac{a^4}{k^2}-1}-\frac32
\mathrm{arccos}\frac{k}{a^2}\right)^{2/3}\ .
\ee
We also use the asymptotic expression for the Airy function occurring
in \eqref{Bessel}, see \cite{AS2},
\be
\mathrm{Ai}\left(\left[\frac{k}{2\hbar}\right]^{2/3}\zeta\right)
\sim \pi^{-1/2}\left[-\left(\frac{k}{2\hbar}\right)^{2/3}\zeta\right]
^{-1/4}\sin\theta_k\ ,
\ee
where
\be
\theta_k= -\frac{k}{3\hbar}\zeta^{3/2}+\frac{\pi}{4}\ .
\ee
The classical trajectory is then recovered through the principle
of constructive interference: We look for the extremum of the phase
\be
\lb{Sk}
S_k\equiv\theta_k\pm \frac{k\phi}{\hbar}
\ee
of the total wave function with respect to $k$.
One then easily finds that the requirement $\partial S_k/\partial k=0$
at $k=\bar{k}$ leads to \eqref{phia1}.
One can thus identify $C_p=\bar{k}$.

What happens for $z^2<1$? As one can easily see from the
corresponding expression in \cite{AS2}, $\zeta<0$ and the Airy function
decays exponentially. This is as expected, since
$a^4/k^2<1$ corresponds to the classically forbidden region.

One can also easily check that $S_k$, Eq. \eqref{Sk}, is a solution of the
Hamilton--Jacobi equation arising from \eqref{constraint}
through the substitutions $\pi_a\to \partial S_k/\partial a$ and
$\pi_{\phi}\to \partial S_k/\partial\phi$.

In the case of the conventional scalar field, one gets a change of sign for
the $k^2$-term in \eqref{Ck}. The solutions for $C_k(a)$ are then the
Bessel functions $J_{ik/2\hbar}(a^2/2\hbar)$ and
$J_{-ik/2\hbar}(a^2/2\hbar)$. Since there are no classically forbidden
regions, both solutions seem to be allowed. It can again easily be checked that
the classical solution \eqref{phia2} follows in the formal limit
`$\hbar\to 0$' from the principle of constructive interference: One gets
the two branches of \eqref{phia2} from the two Bessel functions.
This suggests to use one {\em or} the other Bessel function if one wants
to avoid interferences (and thus non-classical behaviour)
at large $a$. Since \eqref{wdw1} is hyperbolic for $\ell=1$, one is free
to impose boundary conditions at constant $a$, that is, one can
either impose one packet or two packets there, depending on whether one wants
one branch of the classical solution to be represented or both.

In the phantom case discussed above, the Wheeler--DeWitt equation
is elliptic; one there only imposes the boundary condition that
$\psi$ goes to zero at $a\to 0$ and that it is at most oscillating at the
other boundaries. This fixes the solution to be $J_{k/2\hbar}(a^2/2\hbar)$
or a superposition thereof.
Explicitly, one would consider the following superposition for the
construction of a wave packet,
\be
\lb{super1}
\psi(a,\phi)=\int_0^{\infty}{\mathrm d}k\ A(k)e^{-ik\phi/\hbar}
J_{k/2\hbar}\left(a^2/2\hbar\right)\ ,
\ee
where $A(k)$ is a function of $k$ that is peaked around a particular
value $\bar{k}$ (e.g. a Gaussian). One would not expect the packet to
exhibit dispersion near the minimum of the classical trajectory,
since the phase of the Bessel function is
not rapidly varying with respect to $k$, in contrast to the case
of a massive scalar field discussed in \cite{Ki88}.
We shall, however, expect the occurrence of a dispersion at large
values of $a$.
We shall discuss this explicitly for the more realistic case
in Sec.~III C below.

Making an analogy to ordinary quantum mechanics, one would
compare the solution in the elliptic case to an `initial
wave function' $\psi(t=0,x)$, whereas the hyperbolic case would
correspond to the time evolution $\psi(t,x)$, since one would have for
\eqref{wdw1} a distinguished set of foliations with respect to
an intrinsic time defined by the scale factor. This intrinsic time
could be used as a physical time with respect to which, for example,
further degrees of freedom could be evolved, cf. Sec.~IV.


\subsection{Quantum phantom cosmology -- exponential phantom potential\label{nolambda}}
For non-zero, exponential potential as in Sec. \ref{class_nolambda}, the
Wheeler--DeWitt equation is most conveniently solved after a
transformation to new variables in such a way that the potential
cancels in front of $\Psi$. This is obtained by first transforming to
light-cone type coordinates $z_1\equiv\alpha+\sqrt{\ell}\phi$,
$z_1\equiv\alpha-\sqrt{\ell}\phi$. For $\ell=1$, these are just the
characteristics of the Wheeler--DeWitt equation. The equation now
takes the form
\be
\hbar^2\frac{\partial^2\Psi}{\partial z_1\partial z_2}+ f\left(z_1,z_2\right)\Psi=0\ ,
\ee
from which a transformation to new variables can be made such that
$f\left(z_1,z_2\right)$ is canceled. This corresponds to a
transformation to the variables

\ben
u_\ell(\alpha,\phi)&=&\frac{\sqrt{2V_0}}{3}
\frac{e^{3\alpha-\frac{\lambda\sqrt{6}}{2}\phi}}{1-\ell\left(\frac{\lambda}{\sqrt{6}}\right)^2}\left(
  \cosh(X)+\frac{1}{\sqrt{\ell}}\frac{\lambda}{\sqrt{6}}\sinh(X)\right)\ , \\
v_\ell(\alpha,\phi)&=&\frac{\sqrt{2V_0}}{3}\frac{e^{3\alpha-\frac{\lambda\sqrt{6}}{2}\phi}}{1-\ell\left(\frac{\lambda}{\sqrt{6}}\right)^2}\left(\frac{1}{\sqrt{\ell}}
  \sinh(X)+\ell\frac{\lambda}{\sqrt{6}}\cosh(X)\right)\ ,
\een
where $X\equiv\sqrt{\ell}(3\phi-\ell\frac{\lambda\sqrt{6}}{2}\alpha)$. For
both the phantom and the ordinary field,
$u_\ell$ and $v_\ell$ are real.
The Wheeler--DeWitt equation in these variables takes the simple form
\be
\hbar^2\left(\frac{\partial^2\Psi}{\partial u_\ell^2}-
\ell\frac{\partial^2\Psi}{\partial v_\ell^2}\right)+\Psi=0\ .
\ee
Making a WKB-approximation ansatz, $\Psi=Ce^{\pm\frac{i}{\hbar}S}$, one
obtains at lowest order the Hamilton--Jacobi equation

\be\label{H-J-eqn}
\left(\frac{\partial S_0}{\partial u_\ell}\right)^2-\ell
\left(\frac{\partial S_0}{\partial v_\ell}\right)^2=1\ .
\ee
This is solved via a separation ansatz by
$S_{0k}=ku_\ell-\sqrt{\ell(k^2-1)}v_\ell$. Of course, the
Hamilton--Jacobi equation is also solved by actions carrying
different signs in front of $u_\ell$ and $v_\ell$. These are obtained
from the one chosen above by rotations in the
$(u_\ell,v_\ell)$-plane. For $\ell=-1$, all solutions can be mapped
onto each other in this way.
This is an obvious consequence of the rotational symmetry
of Eq. (\ref{H-J-eqn}) for $\ell=-1$.
As $u_1>0$ (recall that $\lambda <\sqrt{6}$ for
$\ell=1$) for the conventional scalar
field, here only two solutions can be mapped onto
each other.

 From the classical action $S_{0k}$, the equations of motion are obtained
via $\frac{\partial S_{0k}}{\partial k}|_{k=\bar{k}}=c$. 
(Note that $S_{0k}$ evaluated at $k=\bar{k}$ is always real.) For the
special case $c=0$ and
\bdm
\bar{k}^2=1/E_{\rm pot}=\left(1-\frac{\ell \lambda^2}{6}\right)^{-1}
\edm
one obtains the classical trajectories
\be
\phi(\alpha)=\ell\frac{\lambda}{\sqrt{6}}\alpha\ ,
\ee
cf. \eqref{phi_alpha}.

Plugging this lowest-order ansatz into the Wheeler--DeWitt equation, one
finds that the equation is already satisfied {\em exactly}. Thus we get the
following exact wave packet solution to the Wheeler--DeWitt equation,
\be
\Psi(u_\ell,v_\ell)=\int dk\,
A(k)\left(C_1e^{\frac{i}{\hbar}(ku_\ell-\sqrt{\ell(k^2-1)}v_\ell)}+C_2e^{-\frac{i}{\hbar}(ku_\ell-\sqrt{\ell(k^2-1)}v_\ell)}\right)\ .
\ee
By construction, the classical trajectories can be recovered from this
equation through the principle of constructive interference. We choose for
the amplitude a Gaussian with width $\sigma$ centered around $\bar k$, 
\bdm
A(k)=\frac{1}{(\sqrt{\pi}\sigma\hbar)^{1/2}}e^{-\frac{(k-\bar{k})^2}
{2\sigma^2\hbar^2}}\ .
\edm
Taking $C_1=C_2$ for definiteness, one obtains wave packets of the form
\be
\label{wavepacket}
\psi(u_\ell,v_\ell)\approx C_1\pi^{1/4}
\sqrt{\frac{2\sigma\hbar}{1-i\sigma^2\hbar S_0^{\prime\prime}}}
\exp\left(\frac{iS_0}{\hbar}-\frac{S_0^{\prime 2}}{2(\sigma^{-2}
-i\hbar S_0^{\prime\prime})}\right) + \mathrm{c.c}\ ,
\ee
where a Taylor expansion of $S_{0k}$ has been carried out around
$\bar k$ (primes denoting derivatives with respect to $k$) and the terms of the order $(k-\bar k)^3$ in  the exponent have been
neglected. (For simplicity, in this expression
$S_{0k}(\bar{k})\equiv S_0$.)
This can be done if the Gaussian is strongly peaked around
$\bar k$, that is, if $\sigma$ is sufficiently small.
Since $S_{0k}'(\bar{k})=0$ gives the classical trajectory, the packet
is peaked around it. 
For the conventional scalar field as well as for the phantom field,
the wave packet thus follows the classical trajectory but {\em spreads} as
$v_\ell^2\to\infty$. This can be recognized from \eqref{wavepacket}, since
the term proportional to $[S_{0k}^{\prime\prime}(\bar k)]^2$
in the width of the Gaussian increases without limit,
\be
S_{0k}^{\prime\prime}(\bar k)=\frac{v_{\ell}}{\left(\ell(\bar k^2-1)\right)^{\frac32}}\ .
\ee
It is even more obvious from the absolute square of the wave packet
(neglecting for simplicity the complex conjugate part in \eqref{wavepacket}),
\be
\vert\psi(u_{\ell},v_{\ell})\vert^2\approx\vert C_1\vert^2\sqrt{\pi}\frac{2\sigma\hbar}
{\sqrt{1+\sigma^4\hbar^2(S_0^{\prime\prime})^2}}
\exp\left(-\frac{S_0^{\prime 2}}{\sigma^{-2}+\sigma^2\hbar^2
(S_0^{\prime\prime})^2}\right)\ .
\ee
The spreading occurs due to the non-trivial dispersion relation, that is,
due to the fact that $S_{0k}$ depends non-linearly on $k$.
The semiclassical approximation is thus not valid
throughout configuration space.

For the phantom field we have $u_{-1}\to-\infty$, $v_{-1}\to\infty$ when
we approach the Big-Rip singularity. This singularity thus lies in
a genuine quantum region. Since for $\ell=-1$ one has
\bdm
v_{\ell}^2 \sim e^{6\alpha-\lambda\sqrt{6}\phi}\equiv e^{6\alpha}
V(\phi)\ ,
\edm
it is obvious that the occurrence of the non-trivial potential is
responsible for the dispersion. 

The Big-Rip singularity is thus `smoothed out' --- when the
wave packets disperse, we can no longer use an approximate time
parameter; time and the classical evolution come to an end, and one
is just left with a stationary quantum state. This corresponds to
quantum gravity effects at very large scales. Hitherto such a case
has only be encountered near the turning point of a classically
recollapsing universe, as a consequence of the demand that the wave function
go to zero for large scale factor \cite{Ki88,KieZeh}.

Due to the fact that $u_{1}>0$ for the conventional scalar field
model, here two inequivalent actions exist. Apart from the wave
packet constructed from the function $S_{0k}=ku_1-\sqrt{k^2-1}v_1$, one gets a
second wave packet constructed from
$S_{0k}=-ku_1-\sqrt{k^2-1}v_1$. Moreover, the entire $(\alpha,\phi)$-plane
is mapped into only one quarter of the $(u_1,v_1)$-plane. One
would therefore require the wave packet to vanish at the
boundary of the physical region. The only solution satisfying this
requirement is naturally the trivial one.
To get a non-trivial solution, one has to lessen the boundary
condition and require $\Psi=0$ only at the origin of the $(u_1,v_1)$-plane.
The fact that the wave packet does not vanish at the $u_1=0$ and
$v_1=0$ line is due to the non-normalizability of the wave packet in
both $\alpha$ and $\phi$, which in turn has its origin in the fact that
the classical trajectory has no turning point.

The implementation of this condition results in a wave
packet which vanishes at the Big-Bang singularity, $\Psi\to 0$ as
$\alpha\to -\infty$, and spreads for large $\alpha$.
The Big-Bang singularity does therefore not exist in the quantum theory.
In the phantom field model, no such restriction occurs due to the fact
that the entire $(u_{-1},v_{-1})$-plane represents 
the entire $(\alpha,\phi)$-plane.
The wave packets for both the phantom and the
ordinary scalar field are depicted in Figure~3.

\begin{figure}[h]
\unitlength1cm
\psfrag{X}{\mbox{$u$}}
\psfrag{Y}{\mbox{$v$}}
\begin{center}
\scalebox{0.7}{\includegraphics[angle=0]{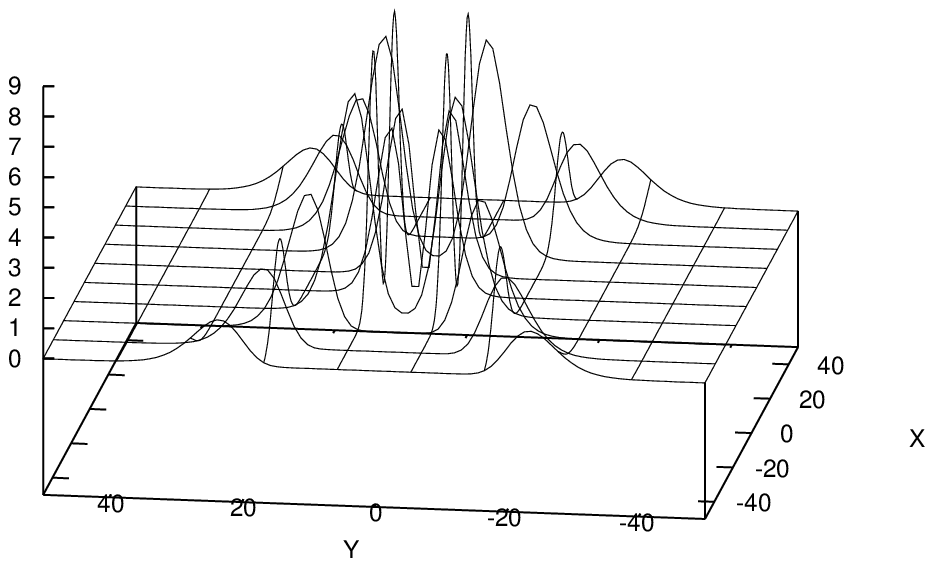}\includegraphics[angle=0]{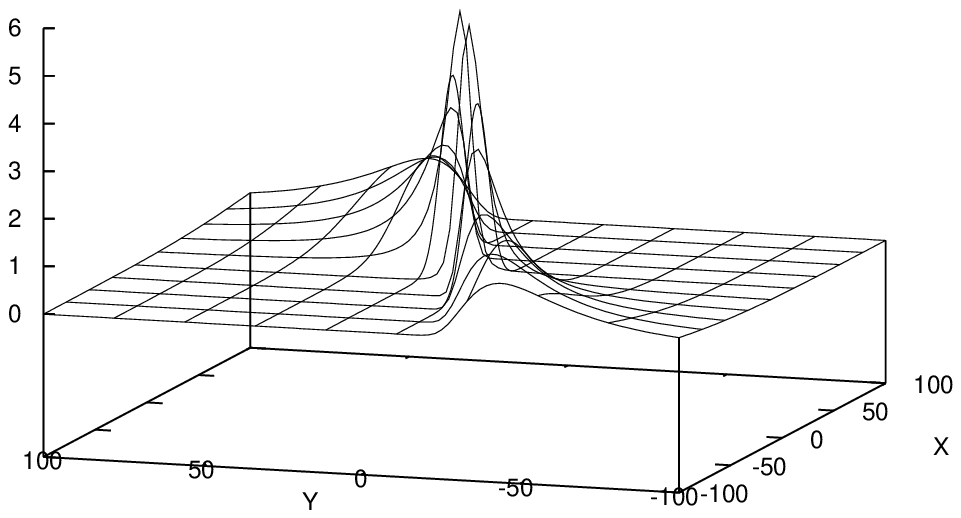}}
\caption{The amplitude of the wave packet \eqref{wavepacket} for an exponential potential solution of the
WDW equation for $\ell=1$ (left) and $\ell=-1$ (right). Here, $\hbar$
was set to unity and parameters $\sigma=0.1$ and
$\lambda={\kappa}/{2}$ have been chosen. The wave packet for the
phantom field model is seen to spread near the classical
singularity. For the scalar field model one has $\Psi\to 0$ at the origin. In
each sector corresponding to one copy of the $(\alpha, \phi)$ plane,
the same wave packet propagates.}
\end{center}
\end{figure}

\subsection{Quantum phantom cosmology --
scalar field fluid and negative cosmological constant}

For the model discussed in Sec. \ref{neglambda}, the classical solutions
require a potential of the form
$V(\phi)=V_0\cosh^2({\phi}/{F})$, cf. \eqref{cosh}.
The Wheeler--DeWitt equation therefore reads
\be
\frac{\hbar^2}{2}\left(\frac{\partial^2}{\partial\alpha^2}-
\ell\frac{\partial^2}{\partial\phi^2}\right)\Psi(\alpha,\phi)+
e^{6\alpha}\left(V_0\cosh^2\left(\frac{\phi}{F}\right)+\frac{\Lambda}{6}\right)\Psi(\alpha,\phi)=0\ .
\ee
The classical singularities lie in a region of large $|\phi|$.
In order to study the quantum behaviour there, it is thus sufficient
to approximate the potential for large $|\phi|$,
\be
\tilde V(\phi)\approx
\frac{V_0}{4}e^{\frac{\pm 2\phi}{F}}\ ,
\ee
where in the following the upper sign refers to positive $\phi$, and the
lower sign to negative $\phi$.
This makes the problem very similar to the one of
Sec. \ref{nolambda}. The Wheeler--DeWitt equation is here simplified
by a transformation on the variables

\ben
u_\ell(\alpha,\phi)&\equiv&\frac{\sqrt{V_0}}{3\sqrt{2}}\frac{e^{3\alpha\pm\frac{\phi}{F}}}{1-\frac{\ell}{9F^2}}\left(\cosh{(X)}\mp\frac{1}{3F\sqrt{\ell}}
\sinh{(X)}\right)\ ,\\
v_\ell(\alpha,\phi)&\equiv&\frac{\sqrt{V_0}}{3\sqrt{2}}\frac{e^{3\alpha\mp\frac{\phi}{F}}}{1-\frac{\ell}{9F^2}}\left(\frac{1}{\sqrt{\ell}}\sinh{(X)}\pm\frac{1}{3F}\cosh{(X)}\right)\ ,
\een
where $X\equiv\sqrt{\ell}\left(3\phi\pm\ell{\alpha}/{F}\right)$.
In these variables, we recover the form
\be
\hbar^2\left(\frac{\partial^2\Psi}{\partial u_\ell^2}-\ell\frac{\partial^2\Psi}{\partial v_\ell^2}\right)+\Psi=0\ .
\ee
Again, one obtains a solution from a WKB ansatz. The Hamilton--Jacobi
equation is again given by (\ref{H-J-eqn}) (this equivalence is, of
course, only formal, since $u_\ell$ and $v_\ell$ are defined
differently). It is again solved by
$S_{0k}=ku_\ell-\sqrt{\ell\left(k^2-1\right)}v_\ell$, where the remarks
of Sec. \ref{nolambda} concerning the choice of action apply here
as well. The equations of motion
obtained for $\frac{\partial S_{0k} }{\partial k}|_{k=\bar k}=0$ are
\be
\phi(\alpha)=\mp\frac{\ell}{\sqrt{3}}\sqrt{\frac{D}{\ell}}\alpha+C_{\bar k,\ell}\ .
\ee
This solution coincides approximately
with the classical solutions (\ref{traj1}):
If one approximates (\ref{traj1}) for
$\ell=-1$ for {\it large} $a$, one gets ($\pm$ label the different
branches of the classical solution)
\be
\phi_\pm(\alpha)=\pm\frac{1}{\sqrt{3}}\sqrt{\frac{D}{\ell}}\alpha\pm
F\ln{\left(2A\right)}~,
\ee
where $\alpha\ge 0$. Therefore, the limit of large positive $\phi$ is
obtained on the $\phi_+$-branch, while the limit
for large negative $\phi$ is reached on the $\phi_-$-branch. On the other hand, an appproximation for {\it small}
$a$ in the case $\ell=1$ yields
\be
\phi_\pm(\alpha)=\pm\frac{1}{\sqrt{3}}\sqrt{\frac{D}{\ell}}\alpha\mp
F\ln{\left(2A\right)}~,
\ee
where $\alpha\le 0$. Due to this, the limit of large positive $\phi$ is
obtained on the $\phi_-$-branch, and for large negative $\phi$ on the
$\phi_+$-branch. Thus the solution to the approximated
Hamilton--Jacobi equation (\ref{H-J-eqn}) coincides with the approximation of
equation (\ref{traj1}). Of course, a special choice for $\bar k$ has to be
made to fix the onset.
The fact that for $\ell=-1$ large $\phi$ correspond to large $a$, and
for $\ell=1$ large $\phi$ correspond to small $a$ is due to
phantom-scalar field duality.

With the help of the classical action $S_{0k}$, the approximate
Wheeler--DeWitt equation can be solved. Again, the WKB ansatz satisfies
the equation exactly. The wave packet is of the same form as in
Sec. \ref{nolambda}, with a different definition of $u_\ell$ and $v_\ell$ and another
choice of the center of the Gaussian, $\bar k$. As in the case of
vanishing cosmological constant, the wave packet spreads
for $v_\ell^2\to\infty$. The Big-Rip singularity in these variables
occurs at $v_{-1}^2\to\infty$, $u_{-1}\to\infty$.
Thus, again, the
singularity is hidden in a quantum regime and the semiclassical
approximation is not valid throughout configuration space.\\
Due to the restriction $D<3$ for the $\ell=1$ model, the same remarks
concerning the range of the coordinates as in Sec.~\ref{nolambda} apply
here. So at the Big Bang, $\Psi\to 0$. 
In analogy to \cite{KieZeh} one would expect quantum effects to
occur also in the region of the classical turning point. This will be
addressed in a future publication.


\section{Discussion and outlook\label{Sect4}}

In our paper we have applied the formalism of standard quantum cosmology
(using the Wheeler--DeWitt equation) to a situation where phantom fields
are present. This is of interest because there are novel features with
regard to both the structure of the equation (elliptic or
ultrahyperbolic instead of hyperbolic) as well as the presence of
new scenarios (Big-Rip singularity at
large scale factors in the classical model). In fact, one of the most
intriguing features is the possible {\it occurrence of quantum effects
for large scale factors}.

For various models we have determined and discussed the
classical trajectories in configuration space. We have then considered
the corresponding Wheeler--DeWitt equations; we have
given various solutions and addressed the
classical limit as well as the behaviour of wave packets
following the classical trajectories in configuration space.
We have found that the packets disperse in the region
of the classical Big-Rip singularity. 
This singularity is thus `smeared out' by quantum effects
at large scale factor. Once the wave packets disperse, no approximate
time parameter can be defined \cite{OUP} and the classical evolution
terminates in a singularity-free way.

For the conventional scalar field model we have found that
the wave packet vanishes at the Big-Bang singularity due to the
implementation of appropriate boundary conditions. In this way, the Big-Bang
singularity is removed from the quantum theory.
This is similar to the avoidance of the singularity in models
of loop quantum cosmology \cite{Bojowald} and shell collapse
\cite{Hajicek}. Without this boundary condition the wave packet would
just have approached the region $\alpha\to-\infty$ without spreading; this lack
of dispersion is a result of the Wheeler--DeWitt equation taking the form of
a free wave equation in this limit.

The present work can be extended in various directions.
The next step would be to add other `conventional' scalar fields
and to investigate the full quantum dynamics. In particular, this would
be of importance for a discussion of the arrow of time
\cite{Zeh}. 
In order to define an appropriate entropy it is necessary
to introduce a set of inhomogeneous degrees of freedom. In the case
of a classically recollapsing universe it has been found that the
arrow of time is correlated with the scale factor of the universe, that is,
the arrow of time must formally reverse at the maximal expansion
\cite{Zeh,KieZeh}.
(The reversal is formal because quantum effects near the classical
turning point do not allow classical observers to survive this region.)
It is of interest to investigate whether a similar behaviour occurs here.
Regarding the classical evolution of the phantom fields depicted on
the left-hand sides of the Figures 1 and 2, one would again
expect a correlation of the entropy with increasing scale factor,
that is, there would be no collapse followed by an expansion but only
two separate branches of expansion separated by a quantum region.
The classical evolution would then start out of the quantum phase
at small scales and describe an expansion of the universe ended
by a quantum phase near the `Big-Rip region'. We shall address this
scenario in a forthcoming paper. 
This will also deal with
a possible quantum phase near weak singularities (sudden future 
singularity, generalized sudden future singularity, type III and type IV) 
as mentioned in the Introduction.

We have based our discussion on the Wheeler--DeWitt equation of
quantum geometrodynamics. More recently, an alternative formulation of
canonical quantum gravity called loop quantum gravity has gained
considerable attention, cf. \cite{OUP}. A major prediction of this approach
is the presence of a discrete structure for geometric operators.
This formalism was applied to cosmology where it led to new
features \cite{Bojowald}. Instead of the usual Wheeler--DeWitt equation
one gets a difference equation for the scale factor. While near the
Planck scale this equation gives different results from the
(differential) Wheeler--DeWitt equation (and therefore can
prevent the occurrence of the classical singularities),
it coincides with it for higher values of $a$. It thus seems that
near the classical Big-Rip singularity the same scenario emerges that
has been discussed in the present paper. However, it is of interest to
investigate quantitatively the differences and similarities of
ordinary quantum phantom cosmology and loop quantum phantom cosmology.
A first paper in this direction has studied the {\em effective}
dynamics from loop quantum cosmology and its consequences
\cite{SST}. The results indicate that the Big Rip can be avoided.
We hope to return to these and related issues in a future publication.


\section*{Acknowledgements}

C.K. is grateful to the Max Planck Institute for Gravitational Physics,
Potsdam, for its kind hospitality while part of this work was done.
He also thanks the University of Szczecin for kind hospitality
during his stay. M.P.D. acknowledges the hospitality of the 
University of Cologne
and the support of the Polish Ministry of Education and Science grant
1PO3B 043 29 (years 2005-2007). B.S. thanks the
Friedrich-Ebert-Stiftung for financial support.
We thank 
Andrei Barvinsky and Alexander Kamenshchik for critical comments.



\end{document}